\documentclass[conference]{IEEEtran}

\usepackage{comment}

\ifCLASSINFOpdf
  \usepackage[pdftex]{graphicx}
   \graphicspath{{../pdf/}{../jpeg/}}
\else
\fi
\IEEEoverridecommandlockouts%
\IEEEpubid{\makebox[\columnwidth]{\hfill 978-1-4799-5433-9/14/\$31.00~\copyright~2014~IEEE} \hspace{\columnsep} \makebox[\columnwidth]{ }}%

\newcommand{\bk}{\mathbf{k}}

\newcommand{\ph}{\varphi}

\newcommand{\ti}{\mathrm{t}}
\newcommand{\nV}{\Psi}
\newcommand{\kbt}{k_B T_L}
\newcommand{\ak}{\alpha_K}
\newcommand{\hw}{\hbar \omega_{p}}
\newcommand{\mass}{m^{*}}
\newcommand{\qe}{\gamma}

\newfont{\iams}{msbm9}
\newcommand{\rind}{\mbox{\iams \symbol{'122}}}
\newcommand{\Itre}{\int_{\scriptstyle \rind^{3}}}
\newcommand{\Ipm}{\int_0^\pi \! \! d\ph' \int_{-1}^1 \! d\mu' \:}
\newcommand{\Iwmp}{\int_{0}^{+ \infty} \! \! dw \int_{-1}^1 \! d\mu
                   \int_0^\pi \! \! d\ph \:}

\newcommand{\argf}{( \mathbf{\cdot} )}
\newcommand{\ot}{\frac{1}{2}}
\newcommand{\dm}{\displaystyle}

\begin{document}
%
\title{Boundary conditions effects by Discontinuous Galerkin Solvers for Boltzmann-Poisson models of Electron Transport}

\author{\IEEEauthorblockN{Jos\'e A. Morales Escalante}
\IEEEauthorblockA{Institute for Computational \\ Engineering and Sciences (ICES) \\
The University of Texas at Austin\\
Austin, Texas 78712-1229\\
Email: jmorales@ices.utexas.edu}
\and
\IEEEauthorblockN{Irene M. Gamba}
\IEEEauthorblockA{Department of Mathematics and ICES \\
The University of Texas at Austin\\
Austin, Texas 78712-1229\\
Email: gamba@math.utexas.edu}
}


%


\maketitle

\begin{abstract}

In this paper we perform, by means of Discontinuous Galerkin (DG) Finite Element Method (FEM) based numerical solvers for 
Boltzmann-Poisson (BP) semiclassical models of  hot electronic transport in semiconductors, 
a numerical study of reflective boundary conditions in the BP system,
such as specular reflection, diffusive reflection, and a mixed convex combination of these reflections, 
and their effect on the behavior of the solution. 
A boundary layer effect is observed in our numerical simulations 
for the kinetic moments related to diffusive and mixed reflection. 


\end{abstract}


%
\IEEEpeerreviewmaketitle

\section{Introduction}

\hfill June 30, 2014 

The dynamics of electron transport in modern semiconductor devices can be described by the semiclassical Boltzmann-Poisson (BP) model:
\begin{equation}
\frac{\partial f_i}{\partial t} + \frac{1}{\hbar} \nabla_{\vec{k}} \,
\varepsilon_{i} (\vec{k}) \cdot \nabla_{\vec{x}} f_i -
 \frac{q_i}{\hbar} \vec{E} \cdot \nabla_{\vec{k}} f_i = \sum_{j} Q_{i,j}
\label{BE}
\end{equation}
\begin{equation}
\nabla_{\vec{x}} \cdot \left( \epsilon \, \nabla_{\vec{x}} V \right)
 = \sum_{i} q_i \rho_{i} - N(\vec{x}), 
 \vec{E} = - \nabla_{\vec{x}} V
  \label{pois}
\end{equation} 
\ \\[-10pt]
$f_{i}(\vec{x},\vec{k},t)$ is the probability density function (pdf) over phase space $(\vec{x},\vec{k})$ of a carrier in the $i$-th energy band in position $\vec{x}$, 
with crystal momentum $\hbar \vec{k}$ at time $t$.  The collision operators $Q_{i,j}(f_i,f_j)$ model 
$i$-th and $j$-th carrier recombinations, collisions with phonons or generation effects.  $\vec{E}(\vec{x},t)$ is the electric field, 
$\varepsilon_{i}(\vec{k})$ is the $i$-th energy band surface, the $i$-th charge density $\rho_{i}(t,\vec{x})$ is the k-average of $f_i$, 
and $N(\vec{x})$ is the doping profile. 

Deterministic solvers for the BP system using Discontinuous Galerkin (DG) FEM have been proposed in \cite{CMAME, iwceCGMS} to model 
electron transport along the conduction band for 1D diodes and 2D double gate MOSFET devices. In \cite{CMAME}, the energy band $\varepsilon(\vec{k})$ model used 
was the nonparabolic Kane band model. These solvers are shown to be competitive with Direct Simulation Monte Carlo (DSMC) methods \cite{CMAME}. The energy band models 
used in \cite{iwceCGMS} were the Kane and Brunetti, $\varepsilon(|\vec{k}|)$ analytical models, but implemented numerically for benchmark tests.

Boundary conditions (BC) for BP models in $(\vec{x},\vec{k})$-boundaries vary according to the considered device and physical situation.
For example, considering electron transport along a single conduction band:

Charge neutrality boundary conditions in 1D and 2D devices are given by:
\begin{equation}
f_{out}(t,\vec{x},\vec{k}) |_{\Gamma} = N_D(\vec{x}) \frac{ f_{in}(t,\vec{x},\vec{k}) |_{\Gamma} }{\rho_{in}(t,\vec{x})}, \quad \vec{x} \in \Gamma \subseteq \partial \Omega_{\vec{x}}
\end{equation}
Specular reflection BC over the Neumann Inflow Boundary \\
$\Gamma_N^- =  \{(\vec{x},\vec{k}) : \vec{x} \in \Gamma_N, \, \vec{k} \in \Omega_{k}, \, \nabla_{\vec{k}} \, \varepsilon(\vec{k}) \cdot \eta(\vec{x}) < 0  \} $,  
with outward unit normal $ \eta(\vec{x})$ (the Neumann boundary $\Gamma_N$ usually defines insulating boundaries) is imposed by:
\begin{equation}
 f^{spec}(\vec{x},\vec{k},t) = f(\vec{x},\vec{k}',t) \quad \mbox{for} \quad (\vec{x},\vec{k}) \in \Gamma_N^-, \quad t>0.
\end{equation}
 for
$
 \vec{k'} \quad \mbox{s.t.} \quad \nabla_{\vec{k}}\varepsilon(\vec{k}') = \nabla_{\vec{k}}\varepsilon(\vec{k}) - 2(\nabla_{\vec{k}}\varepsilon(\vec{k}) \cdot \eta(\vec{x}))\eta(\vec{x}) .
$

Diffusive reflection BC is known in the kinetic theory of gas dynamics. The distribution function at the Inflow boundary is proportional
to a Maxwellian \cite{Jungel}. For $(\vec{x},\vec{k}) \in \Gamma_N^-$:
\begin{equation}
f^{diff}(\vec{x},\vec{k},t)=Ce^{-\varepsilon(\vec{k})/K_B T} \int_{\nabla_{\vec{k}}\varepsilon(\vec{k}) \cdot \eta > 0} \nabla_{\vec{k}}\varepsilon(\vec{k}) \cdot\eta(\vec{x})f d\vec{k}
\end{equation}

Mixed reflection BC models the reflection of the electrons from a rough boundary, 
giving by the reflected wave for convex combination of  specular and  diffuse components

$
f(\vec{x},\vec{k}) = p \, f^{spec}(\vec{x},\vec{k}) + (1-p) f^{diff}(\vec{x},\vec{k}), \quad (\vec{x},\vec{k}) \in \Gamma_N^-
$

where the probability $p$ is sometimes called the specularity parameter. It can either be constant, or be a function of the momentum $p(\vec{k})$, as in \cite{Soffer}.

\subsection{BP system with $\vec{k}$ coordinate 
transformation assuming a Kane Energy Band}

The Kane Energy Band Model is a dispersion relation between the conduction energy band $\varepsilon$ (measured from a local minimum)
and the norm of the electron wave vector $|k|$, given by the analytical function ($\alpha$ is a constant parameter,
$m^*$ is the electron reduced mass for Si, and $\hbar$ is Planck's constant):

\begin{equation}
\varepsilon (1 + \alpha \varepsilon) = \frac{\hbar^2 |k|^2}{2m^*}  
\end{equation}

For our preliminary numerical studies we will use a Boltzmann-Poisson model as in \cite{CMAME},
in which the conduction energy band is assumed to be given by a Kane model. 

We use the following dimensionalized variables, with the related characteristic parameters:

{
$
\dm
t = {\ti}/{t_*} , 
 (x,y) = {\vec{x}}/{\ell_*}, 
\ell_* = 10^{-6} m$, $t_* = 10^{-12} s$,  $V_* = 1 \mbox{V}
$
}

A transformed Boltzmann transport equation is used as in \cite{CMAME} as well,
where the coordinates used to describe $\vec{k}$ are:  $\mu$, the cosine of the polar angle, the azimuthal angle $\varphi$,
and the dimensionless Kane Energy $ w = {\varepsilon}/{K_B T} $ 
($K_B$ is Boltzmann's constant, $T$ is the lattice temperature, and $ \ak = \alpha {K_B T} $):

\begin{equation}
\vec{k} = \vec{k}(w,\mu,\varphi) =
\end{equation}
$
\frac{\sqrt{2 m^* \kbt  }}{\hbar} \sqrt{w(1+\ak w)} 
\left(
\mu,\sqrt{1-\mu^2} \cos \ph, \sqrt{1-\mu^2} \sin \ph  
\right)
$

A new unknown function $\Phi$ is used in the transformed Boltzmann Eq. \cite{CMAME}, which is proportional to the Jacobian of the transformation and to the density of states:

\begin{equation}
\Phi(t, x, y, w, \mu, \ph) = s(w) f(\ti, \vec{x}, \vec{k}) \, ,
\end{equation}

where
$
 s(w) = \sqrt{w(1+\ak w)}(1+2\ak w).
$

The transformed Boltzmann transport equation for $\Phi$ in \cite{CMAME} is:

\begin{equation}
{\partial_t \Phi} 
+ {\nabla_{\left(x,y,w,\mu,\ph \right)}} \cdot \left( \Phi \vec{g} \right) 
= C(\Phi)
\label{eqPhi}
\end{equation}
Regarding $\vec{g} = (g_1,g_2,g_3,g_4,g_5)$, the functions $g_i$, for  $i=1,2$ are proportional to the $k_x, \, k_y$ cartesian components of the electron group velocity 
$\frac{\partial w}{\partial \vec{k}} $ written as functions of the coordinates $w$, $\mu$, $\ph$. 
The functions $g_i$, for $i=3,4,5$, represent the transport in $\vec{k}$-space due to the electric field, time and position dependent.

The right hand side of (\ref{eqPhi}) is the collision 
operator, after having applied the Fermi Golden Rule  for electron-phonon scattering,
that depends on the energy differences between transition states, 
{
\begin{eqnarray*}
&&C(\Phi)(t,x,y,w,\mu,\ph) =  \\
 && \ s(w) \left\{ c_{0} \Ipm
\Phi(t,x,y,w,\mu ',\ph')
   \right. \\
&& \left. \ \ +  \Ipm [ c_{+} \Phi(t,x,y,w + \qe,\mu ',\ph')\right.\\
  && \left.  \ \ \qquad\qquad\qquad+ c_{-} \Phi(t,x,y,w - \qe,\mu ',\ph') ] \right\}
\\[5pt]
&& - 2 \pi [c_{_0} s(w) +  c_{_+} s(w - \qe) + c_{_-} s(w + \qe)]
   \Phi(t,x,y,w,\mu,\ph) ,
\end{eqnarray*}
}
with the dimensionless parameters $\qe = \frac{\hw}{\kbt},$
{
\begin{equation}
\dm
  (c_0, c_+, c_-) = \frac{2 \mass \, t_*}{\hbar^3} \sqrt{2 \, \mass \, \kbt}
                    \left(K_0 , (n_{q} + 1) K , n_{q} K \right) ,
\end{equation}
}


The electron density is:
{
$$
  \dm n(t_* t, \ell_* x, \ell_* y)
   = \Itre f \: d \bk
   = \left( \frac{\sqrt{2 \,\mass \kbt }}{\hbar} \right)^{\! \! 3}
   \rho(t,x,y) \, ,
$$
}
where
\begin{equation}
  \rho(t,x,y) = \Iwmp \Phi (t,x,y,w,\mu,\ph) \, .
\label{dens}
\end{equation}
Hence, the dimensionless Poisson equation is
\begin{equation}
\label{pois}
 \frac{\partial}{\partial x} \left( \epsilon_{r} \frac{\partial \nV}{\partial x}
 \right)
 +
 \frac{\partial}{\partial y} \left( \epsilon_{r} \frac{\partial \nV}{\partial y}
 \right)
 = c_{p} \left[ \rho(t,x,y) - \mathcal{N}_{D}(x,y) \right]
\end{equation}
%

$  \mathcal{N}_{D}(x,y) =
\left( \frac{\sqrt{2 \,\mass \kbt }}{\hbar} \right)^{\! \! -3}
N_{D}(\ell_* x, \ell_* y), \,$\\
$
c_p = \left( \frac{\sqrt{2 \,\mass \kbt }}{\hbar} \right)^{\! \! 3}
\frac{\ell_*^{2} q}{\epsilon_{0}} .
$

\section{Numerics: Discontinuous Galerkin Method for BP and Boundary Conditions Implementation}

\subsection{DG Method Formulation}

The DG Method formulation for the transformed Boltzmann Eq. that we consider in this work was developed in \cite{CMAME},
to which we refer for more details. We summarize the basics of the formulation below.

\subsubsection{Domain - 2d-$\vec{x}$ Device, 3d-$\vec{k}$ Space}

The domain of the devices to be considered 
can be represented by means of a 
{rectangular grid} 
in both position and momentum space, i.e.:

$i=1: N_x, j=1 : N_y, k=1 : N_w, m=1 : N_\mu, n=1 : N_\ph$,

$
\Omega_{I}=\left[ x_{i - \ot} ,  x_{i + \ot} \right] \times
\left[ y_{j - \ot} ,  y_{j + \ot} \right] \times
\left[ w_{k - \ot} ,  w_{k + \ot} \right] \times
\left[ \mu_{m - \ot} ,  \mu_{m + \ot} \right] \times
\left[ \ph_{n - \ot} ,  \ph_{n + \ot} \right]
$


$
 x_{i \pm \ot} = x_{i} \pm \frac{\Delta x_{i}}{2} \, , \quad
 y_{j \pm \ot} = y_{j} \pm \frac{\Delta y_{j}}{2}\, ,  \quad
 w_{k \pm \ot} = w_{k} \pm \frac{\Delta w_{k}}{2}\,
$

$
 \mu_{m \pm \ot} = \mu_{m} \pm \frac{\Delta \mu_{m}}{2}\, , \quad
 \ph_{n \pm \ot} = \ph_{n} \pm \frac{\Delta \ph_{n}}{2}.
$



$\Phi_h$ will denote the Piecewise Linear Approximation of $\Phi$ 
in a given cell $\mathring{\Omega}_{I}$, with the multi-index $I=ijkmn$: 

$
\Phi_h \, = \,
T_{ijkmn}(t) + 
X_{ijkmn}(t) \, \frac{(x - x_{i})}{\Delta x_{i}/2} +
Y_{ijkmn}(t) \, \frac{(y - y_{j})}{\Delta y_{j}/2} + \\
W_{ijkmn}(t) \, \frac{(w - w_{k})}{\Delta w_{k}/2} +
M_{ijkmn}(t) \, \frac{(\mu - \mu_{m})}{\Delta \mu_{m}/2} +
P_{ijkmn}(t) \, \frac{(\varphi - \varphi_{n})}{\Delta \varphi_{n}/2} 
$ 

\subsubsection{Discontinous Galerkin (DG) Formulation for the
Transformed Boltzmann - Poisson (BP) System}

On a cartesian grid, for each element $\Omega_I$, find $\Phi_h$ in $V_h$ (piecewise linear polynomial space) s.t. for any test function $v_h \in V_h$
{
\begin{eqnarray*}\label{boltztrans}
\hspace{-30pt}  &&\dm \int_{\Omega_I} \frac{\partial \Phi_h}{\partial t} v_h d\Omega - \int_{\Omega_I} \frac{\partial {v_h }}{\partial x} \left( g_{1} \, \Phi_h \right) d\Omega   \nonumber \\
 && \hspace{-30pt} \mbox{ } - \int_{\Omega_I} \frac{\partial {v_h}}{\partial y} \left( g_{2} \, \Phi_h \right) d\Omega  - \int_{\Omega_I} \frac{\partial v_h}{\partial w} \left( g_{3} \, \Phi_h \right) d\Omega  \nonumber \\
 && \hspace{-30pt} \mbox{ } - \int_{\Omega_I} \frac{\partial v_h}{\partial \mu} \left( g_{4} \, \Phi_h \right) d\Omega 
- \int_{\Omega_I} \frac{\partial v_h}{\partial \varphi} \left(g_{5} \, \Phi_h \right) d\Omega \nonumber \\ 
&& \hspace{-30pt} \mbox{ }  + F_{x}^{+} - F_{x}^{-} + F_{y}^{+} - F_{y}^{-}  +  F_{w}^{+} - F_{w}^{-} + F_{\mu}^{+} - F_{\mu}^{-}   \nonumber \\
\hspace{-30pt} && \mbox{ }  + F_{\varphi}^{+} - F_{\varphi}^{-} = \int_{\Omega_I} C(\Phi_h) v_h d\Omega,
\end{eqnarray*}
}

$F^{\pm}$'s denote boundary integrals, for which
the value of $\Phi$ at the boundary is given by the Numerical Upwind Flux rule.



\subsubsection{Algorithm for DG-BP, from $t^{n}$ to $t^{n+1}$}
\ \\
{ (Dynamic Extension of Gummel Iteration Map) }

1.- Compute electron density {$\rho$}, use it to... \\
2.- Solve Poisson Eq. (by Local DG)
for the potential, then get the electric field {$\vec{E}$}. Compute then {$g_i$}'s transport terms. \\
3.- Solve by DG the transport part of Boltzmann Equation. Method of lines (ODE system) for the time-dependent coefficients of {$\Phi_h$} (degrees of freedom) obtained. \\
4.-Evolve ODE system by time stepping from {$t^{n}$} to { $t^{n+1}$}. \\
(If partial time step necessary, repeat Step 1 to 3 as needed).

\subsection{Numerical Implementation of Reflection Boundary Conditions (BC) by DG schemes}

\subsubsection{Specular Reflection BC}

Specular reflection at boundaries $y=0, \, L_y$ is expressed in angular coordinates by:

{
$
\Phi^{spec}_h(t,x,y,w,\mu,\varphi) = \Phi_h(t,x,-y,w,\mu,\pi - \varphi) \,
$
} 


Defining  $n' = N_{\varphi} -n + 1 $,
$\, I = i0kmn, \, I' = i1kmn'$, \\
if $(x,y,w,\mu,\varphi) \in \Omega_{I}$, then $(x,-y,w,\mu,\pi - \varphi) \in \Omega_{I'}$.
This implies that the $\Phi_h$ coefficients satisfy, taking $ \Delta \varphi_{n'} = \Delta \varphi_n $ :

$
T_I = T_{I'}, \quad  X_I  =  X_{I'}, \quad Y_I = - Y_{I'}, \\
W_I = W_{I'}, \quad  M_I  =  M_{I'}, \quad P_I = - P_{I'}.
$



\subsubsection{Diffusive Reflection BC} 

We define the DG  approximate  diffusive function $\Phi_h^{diff} \in V_h^1$ as follows:  

{
\noindent Use the projection $\Phi_h :=\Pi \Phi(t,x,y,w,\mu,\varphi) \in V_h^1$  and set 

$\sigma_h (x,y,t):=
\int_{ \pm \cos \varphi \geq 0  } |g_2| \, \Phi_h \, dw d\mu d\varphi 
\, \in V_h^1.
$
Next, since $ \Phi^{diff} = C \, \sigma_h \, e^{-w} s(w)$, then set  the approximate by projecting  $\Phi_h^{diff} := \Pi \Phi^{diff} \in V_h^1.
$\\}

\subsubsection{Mixed Reflection BC}  These  conditions are numerically approximated by taking the convex combination: \\
{
$
\Phi^{mr}_h = p(\vec{k}) \, \Phi_h^{spec} + \big(1-p(\vec{k})\big) \Phi_h^{diff}
\,
$
}
of specular and diffusive reflections respectively, where the
we use the momentum-dependent specularity parameter \cite{Soffer} given by the probability 
$p(\vec{k}) = e^{-4 L_r^2 |k|^2 \cos^2 \Theta}  = \exp(-4 l_r^2 w(1+\ak w) \sin^2\varphi) = p(w,\varphi),
$  with
$l_r$ the (normalized) \emph{rms} rough interface height \cite{rms_hight}. 

\section{Preliminary Numerical Results}

In our preliminary numerical simulations 
we consider a 2D n bulk Silicon with rectangular geometry in $(x,y)$
(width: $L_x = 0.15\mu m$, height: $L_y = 12 nm$)
to completely isolate the effect of the reflective boundary conditions on the kinetic moments for this benchmark case. 
The respective domain in $\vec{k}(w,\mu,\varphi)$ is rectangular in 3D. \\

Initial Condition: {$ \left. \Phi(w) \right|_{t=0} \propto N e^{-w}s(w)$}. Final Time: 1.0ps\\
{Boundary Conditions} in {$\vec{k}$}-space: { a cut-off is set at  $w=w_{max}$ with
{ $\Phi$}  machine zero.} \\
This is the only BC needed in $\vec{k}(w,\mu,\varphi)$, since the transport normal to the boundary is analitically zero 
at boundaries related to the following 'singular points': \\
At {$w=0$, $g_3 = 0$}. 
At {$\mu = \pm 1$, $g_4$ = 0.} 
At {$\varphi = 0, \pi$, $g_5$ = 0.}\\
BC in {$\vec{x}$}-space: we set neutral charges at boundaries\\ $x=0,\, x=0.15 \mu m$.\\
The Potential-bias BC is set as either:\\  
$\Delta V = \left. V \right|_{x=0.15 \mu m} - \left. V \right|_{x=0} = 0.5, 1.0, \,$ or $\, 1.5$ Volts.\\
The reflection BC, either specular, diffusive or mixed,  are set at  $y=0, y=12 nm$. \\
The number of cells used in the simulation were: \\
$N_x = 24, \, N_y = 12, \, N_w = 30, \, N_{\mu} = 8, N_{\varphi} = 6$.

We present plots of the Average Energy $e$ and Momentum $U$ vs. Position $(x,y)$
at the final time of $t=1.0$ps with a $\Delta V = 1.0$ Volt bias
for the different specular, diffusive and mixed reflection BC implemented.
A boundary layer was observed in the plots of 
the average density, average energy, and average momentum 
for the diffusive and mixed reflection cases
in the boundaries where these reflection conditions are applied,
compared to the specular case in which these moments are constant w.r.t. position 
for the benchmark case considered.
Boundary layers were also observed for the biases of $\Delta V = 0.5, 1.5$ Volts,
obtaining higher values for average energy and momentum when increasing the bias as expected.
A point to mention is that a DSMC solver for BP 
would have a hard time to
resolve the details of the momentum to the scales present in the momentum plots for our deterministic solver. 

\begin{figure}
 \centering
\includegraphics[angle=0,width=0.65\linewidth]{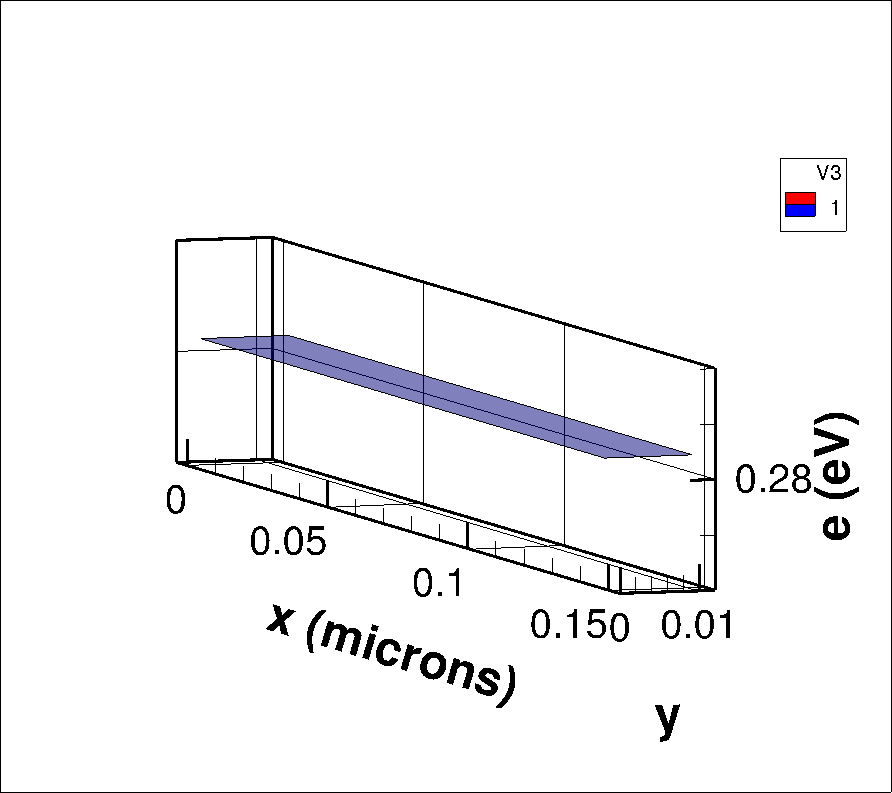}
\end{figure}

\begin{figure}
\centering
\includegraphics[angle=0,width=0.65\linewidth]{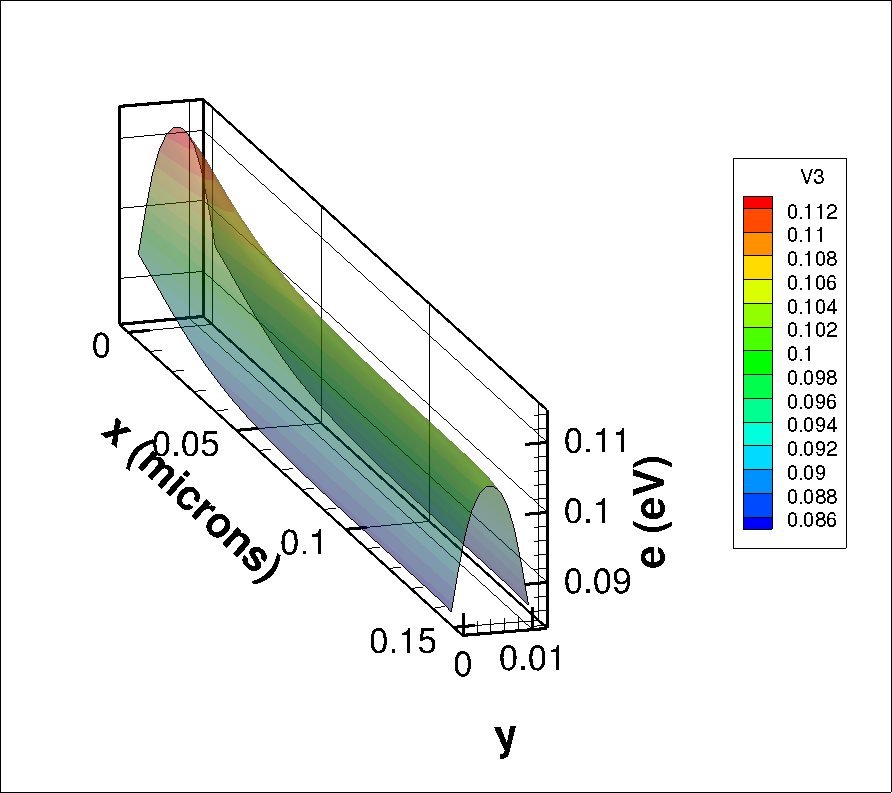}
\caption{  Mean energy $e$ (eV) vs. Position $(x,y)$ in $(\mu m)$  plots for Specular (top) and Diffusive (right above) Reflection.} 
\end{figure}

\begin{figure}
\centering
\includegraphics[angle=0,width=0.65\linewidth]{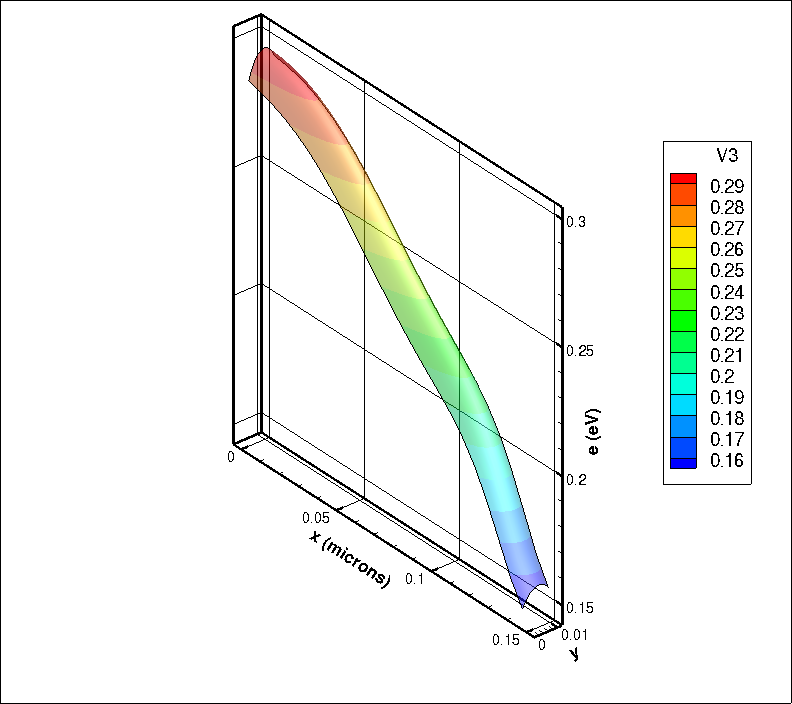}
\end{figure}

\begin{figure}
\centering
\includegraphics[angle=0,width=0.65\linewidth]{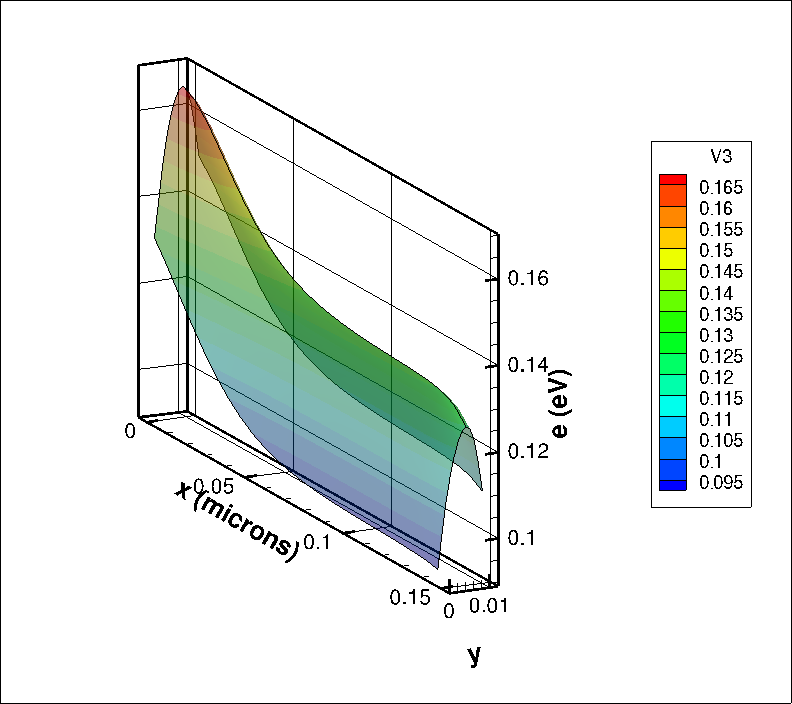}
\caption{  Mean energy $e$ (eV) vs. Position $(x,y)$ in $(\mu m)$  plots for Mixed $p(\vec{k})$, $l_r=0.1$ (top left) \& Mixed $p(\vec{k})$, $l_r = 0.5$ (above) Reflection.} 
\end{figure}


\begin{figure}
\centering
\includegraphics[angle=0,width=0.65\linewidth]{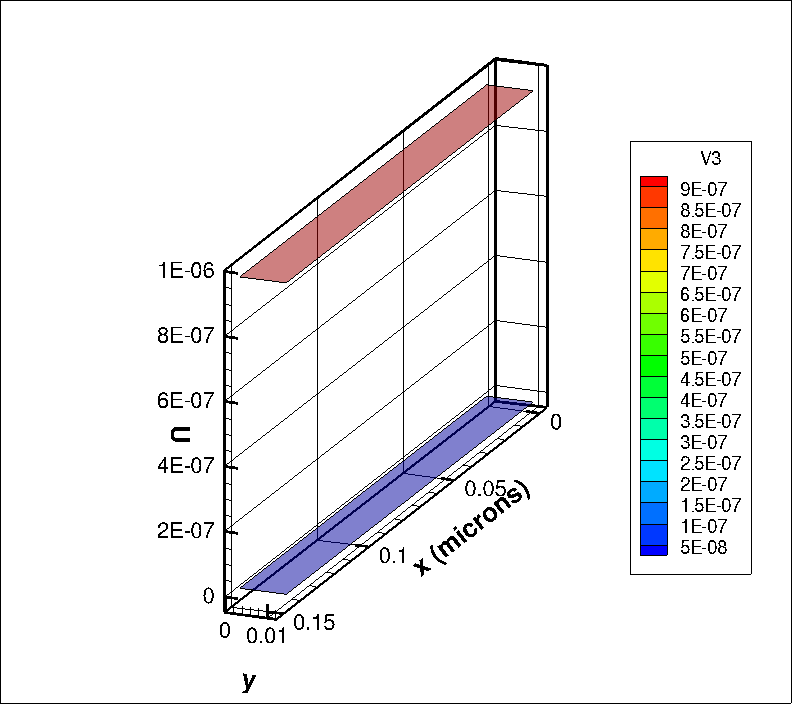} 
\end{figure}

\begin{figure}
\centering
\includegraphics[angle=0,width=0.65\linewidth]{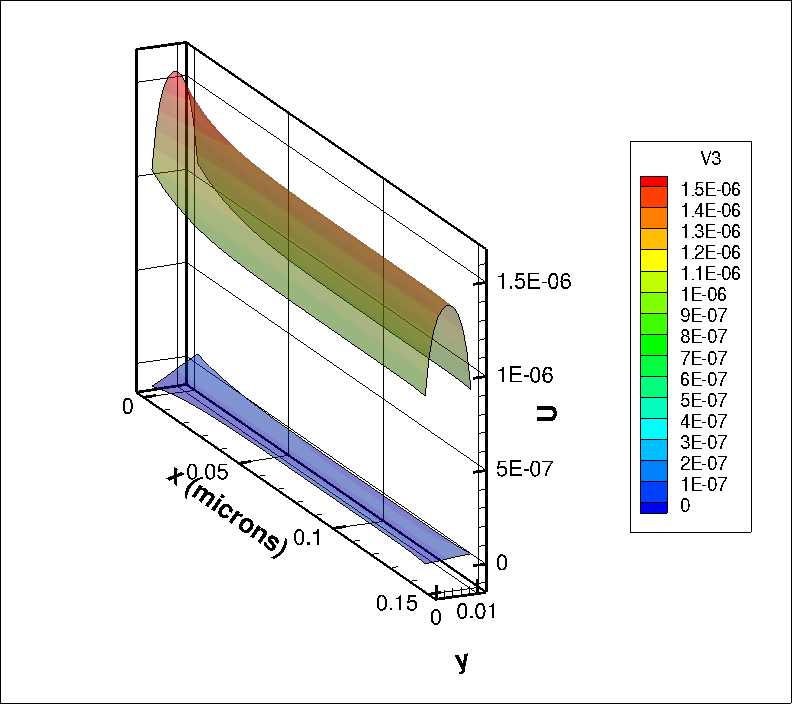} 
\end{figure}

\begin{figure}
\centering
\includegraphics[angle=0,width=0.65\linewidth]{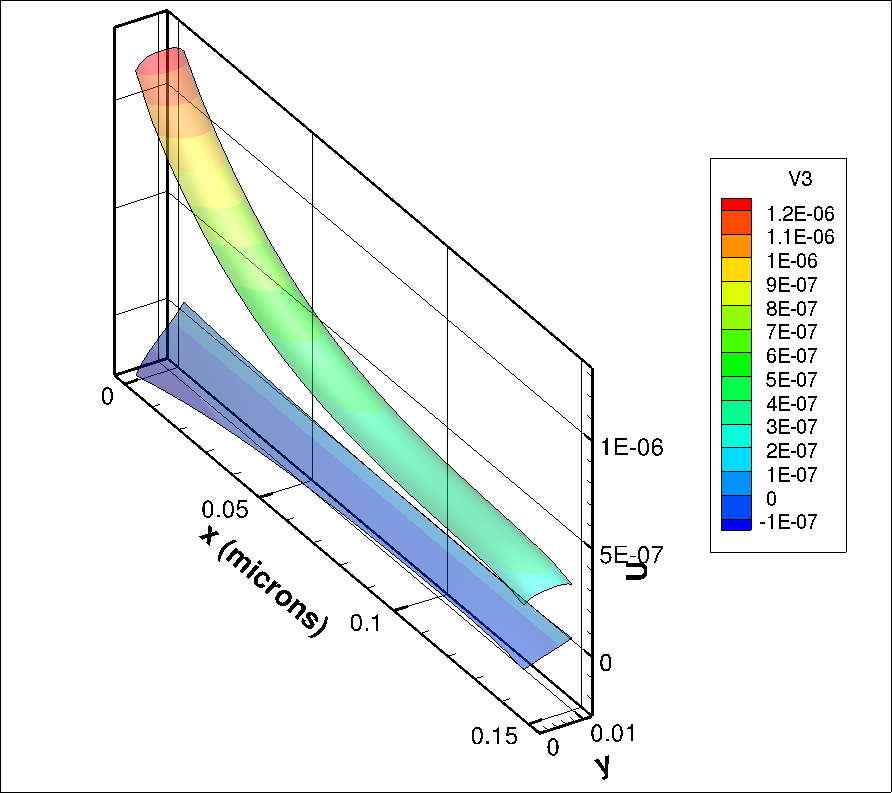} 
\end{figure}

\begin{figure}
\centering
\includegraphics[angle=0,width=0.65\linewidth]{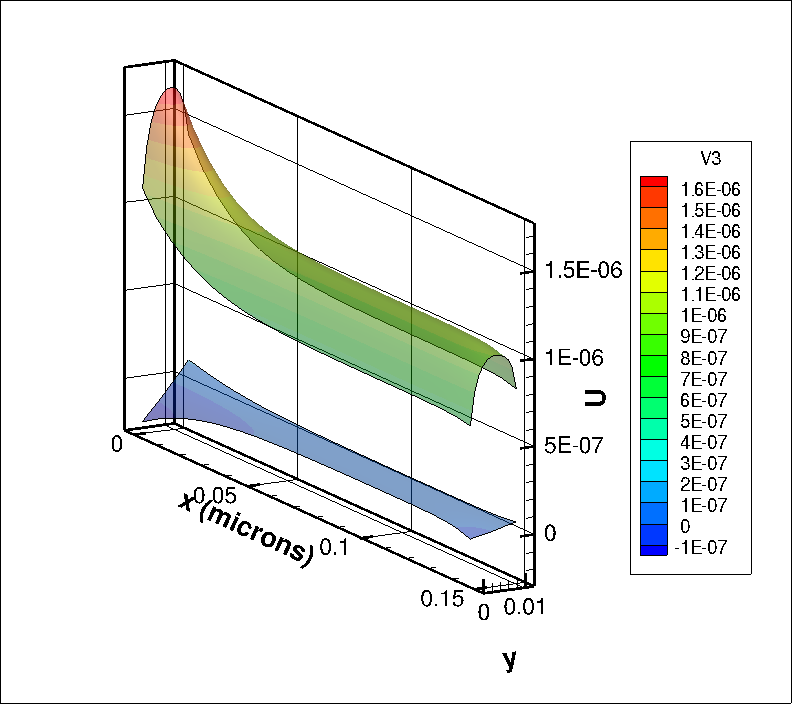} 
\caption{ Momentum $U (10^{28} \frac{cm^{-2}}{s})$ vs. Position $(x,y)$ in $(\mu m)$  for Specular (left below), Diffusive (bottom left), Mixed $p(\vec{k})$, $l_r=0.1$ (top right) \& Mixed $p(\vec{k})$, $l_r = 0.5$ (right above) Reflection} 
\end{figure}

\section{Conclusion}

A Boundary Layer effect was observed in the Kinetic Moments related to the Diffusive and Mixed Reflection cases. 
Work in Progress is related to the case of a 2D double gate MOSFET device. An extended version with more details and results will be presented \cite{extendVers}.
Future work will consider a study of reflective BC on DG solvers where an EPM full band is numerically implemented for 2D devices in $\vec{x}$. 

\section*{Acknowledgment}
The authors have been partially funded by NSF grants CHE-0934450, DMS-1109625, and DMS-RNMS-1107465.
The first author was funded by a NIMS fellowship given by ICES, U.Texas-Austin.



%

\end{document}